\documentclass{elsarticle}
\biboptions{sort&compress}
\usepackage{graphicx}
\usepackage{times}
\usepackage{amsmath}

\newcommand{\be}{\begin{equation}}
\newcommand{\ee}{\end{equation}}
\newcommand{\ba}{\begin{eqnarray}}
\newcommand{\ea}{\end{eqnarray}}

\begin{document}

%\begin{frontmatter}

\title{Lattice Effective Field Theory for Medium-Mass Nuclei}

\author[a1]{Timo~A.~L\"{a}hde\corref{cor1}}
\ead{t.laehde@fz-juelich.de}
\cortext[cor1]{Corresponding author}

\address[a1]{Institute~for~Advanced~Simulation, Institut~f\"{u}r~Kernphysik, and
J\"{u}lich~Center~for~Hadron~Physics,~Forschungszentrum~J\"{u}lich,
D-52425~J\"{u}lich, Germany}

\author[a2]{Evgeny Epelbaum}
\address[a2]{Institut~f\"{u}r~Theoretische~Physik~II,~Ruhr-Universit\"{a}t~Bochum,
D-44870~Bochum,~Germany}

\author[a2]{Hermann~Krebs}

\author[a3]{Dean~Lee}
\address[a3]{Department~of~Physics, North~Carolina~State~University, Raleigh, 
NC~27695, USA}

\author[a1,a5,a6]{Ulf-G.~Mei{\ss }ner}

\address[a5]{Helmholtz-Institut f\"ur Strahlen- und
             Kernphysik and Bethe Center for Theoretical Physics, \\
             Universit\"at Bonn,  D-53115 Bonn, Germany}
\address[a6]{JARA~-~High~Performance~Computing, Forschungszentrum~J\"{u}lich, 
D-52425 J\"{u}lich,~Germany}

\author[a7]{Gautam~Rupak}
\address[a7]{Department~of~Physics~and~Astronomy, Mississippi~State~University, Mississippi State, MS~39762, USA}

\begin{abstract}
\noindent
We extend Nuclear Lattice Effective Field Theory (NLEFT) to medium-mass nuclei, and present results for the ground states of alpha nuclei 
from $^{4}$He to $^{28}$Si, calculated up to next-to-next-to-leading order (NNLO) in the EFT expansion. This computational advance is made 
possible by extrapolations of lattice data using multiple initial and final states. For our soft two-nucleon interaction, we find that the overall 
contribution from multi-nucleon forces must change sign from attractive to repulsive with increasing nucleon number. This effect is not 
produced by three-nucleon forces at NNLO, but it can be approximated by an effective four-nucleon interaction. We discuss the 
convergence of the EFT expansion and the broad significance of our findings for future \textit{ab initio} calculations.
\end{abstract}


\maketitle  

\section{Introduction}

Several \textit{ab initio} methods are being used to study nuclear structure. These include coupled-cluster 
expansions~\cite{Hagen:2012fb}, the no-core shell model~\cite{Jurgenson:2013yya,Roth:2011ar}, the in-medium similarity renormalization 
group approach~\cite{Hergert:2012nb}, self-consistent Green's functions~\cite{Soma:2012zd}, and 
Green's function Monte Carlo~\cite{Lovato:2013cua}. The use of soft chiral nuclear EFT interactions has stimulated much of the recent progress in \textit{ab initio}
nuclear structure calculations. By ``soft'' interactions, we refer to the absence of strong repulsive forces at short distances. In this letter, we 
address a central question in nuclear structure theory: How large a nucleus can be calculated from first principles using the framework of 
chiral nuclear EFT, and what are the remaining challenges?   

We address this question by using Nuclear Lattice Effective Field Theory (NLEFT) 
to calculate the ground states of alpha nuclei from $^{4}$He to $^{28}$Si.  NLEFT is an \textit{ab initio} method where chiral nuclear EFT is combined 
with Auxiliary-Field Quantum Monte Carlo (AFQMC) lattice calculations.  NLEFT differs from other \textit{ab initio} methods in that it is an unconstrained 
Monte Carlo calculation, which does not require truncated basis expansions, many-body perturbation theory, or any constraint on the nuclear wave 
function. Our NLEFT results are thus truly unbiased Monte Carlo calculations. The results presented here form an important benchmark for 
\textit{ab initio} calculations of larger nuclei using chiral nuclear EFT. Any deficiencies are indicative of shortcomings in the specific nuclear 
interactions, rather than of errors generated by the computational method. Such a definitive analysis would be difficult to achieve
using other methods.   

The lattice formulation of chiral nuclear EFT is described in Ref.~\cite{Borasoy:2006qn}, a review of lattice EFT methods can be found in
Ref.~\cite{Dean_QMC}, and Refs.~\cite{Epelbaum:2008ga,Machleidt:2011zz} provide a comprehensive overview of chiral nuclear EFT.
We have recently applied NLEFT to describe the structure of the Hoyle state~\cite{Epelbaum:2011md,Epelbaum:2012qn} 
and the dependence of the triple-alpha process on the fundamental parameters of nature~\cite{Epelbaum:2012iu}.
These studies show that NLEFT is successful up to $A \simeq 12$ nucleons. In this letter, we report the first NLEFT results for medium-mass nuclei. 
We compute the ground state energies for all nuclei in the alpha ladder up to $^{28}$Si using the lattice action established in 
Refs.~\cite{Epelbaum:2012qn,Epelbaum:2011md,Epelbaum:2009zsa}. 

\section{Chiral nuclear EFT for medium-mass nuclei}

According to chiral nuclear EFT, our calculations are organized in powers of a
generic soft scale $Q$ associated with factors of momenta and the pion mass. 
We label the $\mathcal{O}(Q^0)$ contributions to the nuclear Hamiltonian 
as leading order (LO), $\mathcal{O}(Q^2)$ as next-to-leading order (NLO), and $\mathcal{O}(Q^3)$
as next-to-next-to-leading order (NNLO). The present calculations are performed up to NNLO.  Our LO lattice Hamiltonian includes a significant part of the NLO and higher-order corrections by making use of smeared
contact interactions~\cite{Borasoy:2006qn,Epelbaum:2009pd,Epelbaum:2009zsa}.  See Ref.~\cite{Epelbaum:2009zsa} for a discussion of the interactions used in this work.  As discussed in Ref.~\cite{Epelbaum:2009zsa}, we are using a low-momentum power counting scheme where there are no additional two-nucleon corrections at NNLO beyond the terms already appearing at NLO.    

The NLEFT calculations reported here are performed with a lattice spacing of $a = 1.97$~fm in a 
periodic cube of length $L = 11.82$~fm. Our trial wave function is denoted
$|\Psi_{A}^\mathrm{init}\rangle$, which is a Slater-determinant state composed of delocalized standing waves in the periodic cube,
with $A$ nucleons and the desired spin and isospin quantum numbers.  For simplicity, we describe our calculations using the language of continuous time evolution. The actual AFQMC calculations use transfer matrices with a temporal lattice spacing of $a_t^{}=1.32$~fm~\cite{Dean_QMC}.  
%$a_t^{}=(150~\rm{MeV})^{-1}$~\cite{Dean_QMC}.   

Before we enter into the main part of the calculation, we make use of a low-energy filter based upon Wigner's SU(4) symmetry, where 
the spin-isospin degrees of freedom of the nucleon are all equivalent as four components of an SU(4) multiplet. Let us define
\begin{equation}
H_\mathrm{SU(4)}^{} \equiv H_\mathrm{free}^{} 
+\frac{1}{2} \, C_\mathrm{SU(4)}^{}\sum_{\vec n,\vec n'} 
{:\rho(\vec n)f(\vec n - \vec n')\,\rho(\vec{n}'):},
\label{H_SU4}
\end{equation}
where $f(\vec n - \vec n')$ is a Gaussian smearing function with width set by the average effective range of the two $S$-wave interaction channels, 
and $\rho$ is the total nucleon density. We then apply the exponential of $H_\mathrm{SU(4)}$ to obtain
\begin{equation}
|\Psi_A^{}(t^\prime_{})\rangle \equiv \exp(-H_\mathrm{SU(4)}^{} t^\prime_{}) |\Psi_{A}^\mathrm{init}\rangle.
\label{trial}
\end{equation}
%
%We will refer to  $|\Psi_A^{}(t^\prime_{})\rangle$ 
which we refer to as a ``trial state''.  This part of the calculation is computationally inexpensive since it only requires a single 
auxiliary field and does not generate any sign oscillations in the Monte Carlo calculation.   

Next, we use the LO Hamiltonian $H_\mathrm{LO}^{}$ to construct the Euclidean-time projection amplitude
\begin{equation}
Z_A^{}(t) \equiv \langle\Psi_A^{}(t^\prime_{})| \exp(-H_\mathrm{LO}^{} t) |\Psi_A^{}(t^\prime_{})\rangle,
\end{equation}
from which we compute the ``transient energy''
\begin{equation}
E_A^{}(t) = -\partial[\ln Z_A^{}(t)]/\partial t.
\label{EAt}
\end{equation}
If the lowest eigenstate of $H_\mathrm{LO}^{}$ that possesses a non-vanishing overlap with 
the trial state $|\Psi_A^{}(t^\prime_{})\rangle$ is denoted $|\Psi_{A,0}^{}\rangle$, the energy $E_{A,0}^{}$ of $|\Psi_{A,0}^{}\rangle$ 
is obtained as the ${t\to\infty}$ limit of $E_A^{}(t)$.

The higher-order corrections to $E_{A,0}^{}$ are evaluated using perturbation theory. We compute
expectation values using
\begin{align}
Z_A^\mathcal{O}(t) \equiv \: &
\langle\Psi_A^{}(t^\prime_{})| \exp(-H_\mathrm{LO}^{} t/2)  \nonumber \\
& \times \mathcal{O} \exp(-H_\mathrm{LO}^{} t/2)  |\Psi_A^{}(t^\prime_{})\rangle,
\label{OP}
\end{align}
for any operator $\mathcal{O}$. Given the ratio
\begin{equation}
X_A^\mathcal{O}(t) = Z_A^\mathcal{O}(t)/Z_A^{}(t),
\label{XAt}
\end{equation} 
the expectation
value of $\mathcal{O}$ for the desired state $|\Psi_{A,0}^{}\rangle$ is again obtained in the 
$t\rightarrow\infty$ limit according to
\begin{equation}
X_{A,0}^\mathcal{O} \equiv \langle\Psi_{A,0}^{}| \mathcal{O} |\Psi_{A,0}^{}\rangle = \lim_{t \to \infty}X_A^\mathcal{O}(t),
\end{equation}
which gives the corrections to $E_{A,0}^{}$ induced by the NLO and NNLO contributions.

The closer $|\Psi_A^{}(t^\prime_{})\rangle$ is to $|\Psi_{A,0}^{}\rangle$, 
the less the required projection time $t$. The trial state can be optimized by adjusting both the SU(4) projection time $t^\prime_{}$ and the strength 
of the coupling $C_\mathrm{SU(4)}^{}$ of $H_\mathrm{SU(4)}$. 
%In this work, we introduce an additional tool to the extrapolation analysis.
Here, we show that the accuracy of the extrapolation $t\to\infty$ can be further improved by simultaneously incorporating data from multiple trial states 
that differ in the choice of $C_\mathrm{SU(4)}^{}$. This approach enables a ``triangulation'' of the asymptotic behavior as 
the common limit of several different functions of $t$. 

\section{Extrapolation in Euclidean time}

The behavior of $Z_A^{}(t)$ and $Z_A^{\mathcal{O}}(t)$ at large $t$ is controlled by the low-energy spectrum of $H_\mathrm{LO}^{}$. 
Let $| E\rangle$ label the eigenstates of $H_\mathrm{LO}^{}$ with energy $E$, and let $\rho_{A}^{}(E)$ denote the density of states for a system
of $A$~nucleons. For simplicity, we omit additional labels needed to distinguish degenerate states.  
We can then express $Z_A^{}(t)$ and $Z_A^{\mathcal{O}}(t)$ in terms of their spectral representations,
\begin{align}
Z_A^{}(t) = & \int dE \: \rho_A^{}(E) \:
\big| \langle E |\Psi_A^{}(t^\prime_{})\rangle\big|^2_{} 
\exp(-Et), \\
Z_A^{\mathcal{O}}(t) = & \int dE\,dE^\prime_{} \, \rho_A^{}(E)\,\rho_A^{}(E^\prime_{}) 
\, \exp(-(E+E^\prime_{})t/2), \nonumber \\ 
& \times
\langle\Psi_A^{}(t^\prime_{})|E\rangle \,
\langle E|\mathcal{O}|E^\prime_{}\rangle \,
\langle E^\prime_{}|\Psi_A^{}(t^\prime_{})\rangle,
\end{align} 
from which the spectral representations of $E_A^{}(t)$ and $X_A^{\mathcal{O}}(t)$ are obtained using Eq.~(\ref{EAt}) and Eq.~(\ref{XAt}), respectively. 
We can approximate these to arbitrary accuracy over any finite range of $t$ by taking $\rho_{A}^{}(E)$ to be a sum of energy delta functions,
\begin{equation}
\rho_{A}^{}(E) \approx \sum_{i=0}^{i_\mathrm{max}}c_{i}\delta(E-E_{A,i}^{}),
\end{equation}
where we use $i_\mathrm{max} = 4$ for the calculation of the $^4$He ground state, and $i_\mathrm{max} = 3$ for $A \geq 8$. These choices
give a good description over the full range of $t$ for all trial states, without introducing too many free parameters. Using AFQMC data for
different values of $C_\mathrm{SU(4)}^{}$, we perform a correlated fit of $E_A^{}(t)$ and $X_A^{\mathcal{O}}(t)$ for all operators $\mathcal{O}$
that contribute to the NLO and NNLO corrections. We find that using 2-6 distinct trial states for each $A$ allows for a much more precise
determination of $E_{A,0}^{}$ and $X_{A,0}^{\mathcal{O}}$ than hitherto possible. In particular, we may ``triangulate'' 
$X_{A,0}^{\mathcal{O}}$ using trial states that correspond to functions $X_A^{\mathcal{O}}(t)$ which converge both from above and below.
%thereby bracketing $X_{A,0}^{\mathcal{O}}$.

As the extent of our MC data in Euclidean time is relatively short, we discuss next the level of confidence that we can attribute to our results. In our ``triangulation''
method, the accuracy and reliability of the extrapolation $t \to \infty$ is increased by means of correlated fits to multiple trial states. We first note that the number 
of Euclidean time steps $N_t^{}$ available for the extrapolation does not decrease drastically with the number of nucleons $A$. This inspires confidence that 
our method, which has by now been successfully applied to the structure, spectrum and electromagnetic properties of $^{16}$O in Ref.~\cite{16O_spectrum} 
should also be applicable to heavier systems. Nevertheless, ``spurious early convergence'' in Euclidean time extrapolations should be carefully guarded 
against. If only one trial state is used, this issue arises much more readily. In our ``triangulation'' method, the extrapolation is very strongly constrained by 
the requirement that all observables for all trial states should be described by the same exponential dependence on Euclidean time. Rapid convergence
in $t$ then translates into a small sensitivity to $C_\mathrm{SU(4)}^{}$ at large values of $t$. It is also encouraging to
note that our new extrapolations are consistent with our earlier results for $^{12}$C in 
Refs.~\cite{Epelbaum:2012qn,Epelbaum:2012iu}, which were computed using delocalized plane-wave as well as alpha-cluster trial wave functions.

\section{Lattice Monte Carlo results \label{MC}}

\begin{figure}[t]
\begin{center}
\includegraphics[width=.24\columnwidth]{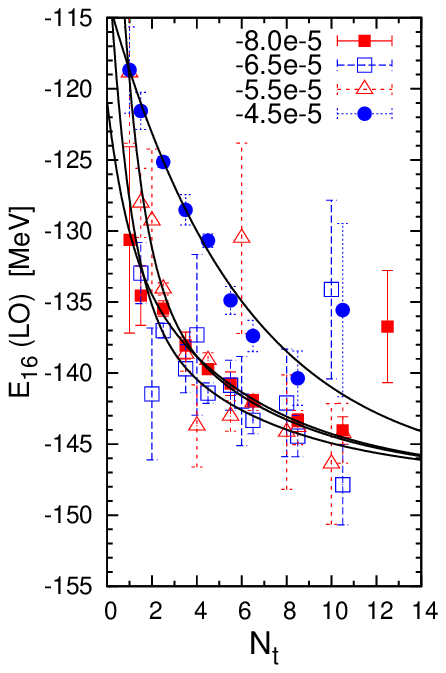}
\includegraphics[width=.24\columnwidth]{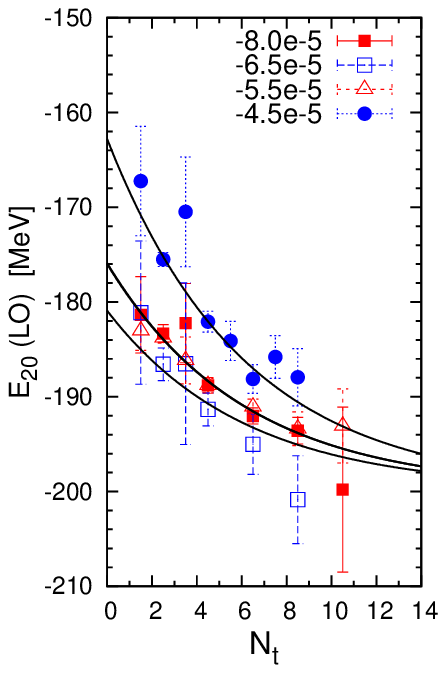}
\includegraphics[width=.24\columnwidth]{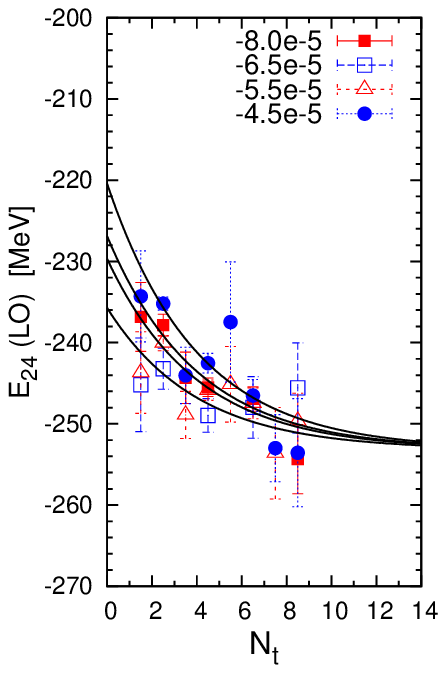}
\includegraphics[width=.24\columnwidth]{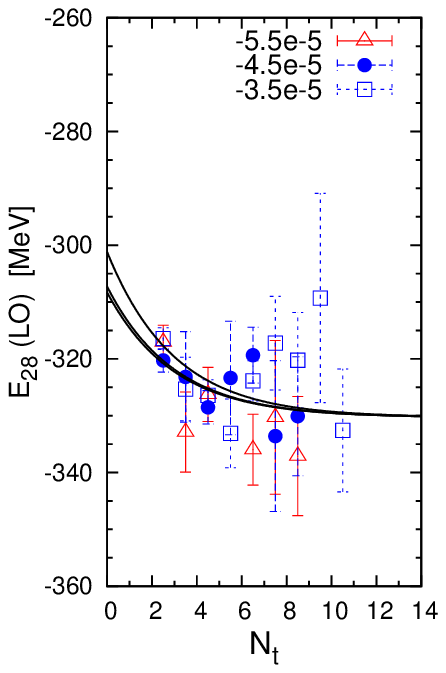}
\end{center}
\vspace{-6mm}
\caption{NLEFT results for the LO transient energy $E_A^{}(t)$ for $A = 16$ to $A = 28$, with $C_\mathrm{SU(4)}^{}$ given 
(in MeV$^{-2}$) for each trial state. The curves show a fit using a spectral density $\rho_A^{}(E)$ given by a sum of three 
energy delta functions. The fits for $E_A^{}(t)$ are correlated with those of Figs.~\ref{fig:NLO1} and~\ref{fig:NLO2}.
\label{fig:LO}}
%\vspace{-6mm}
\end{figure}

\begin{figure}[t]
\begin{center}
\includegraphics[width=.7\columnwidth]{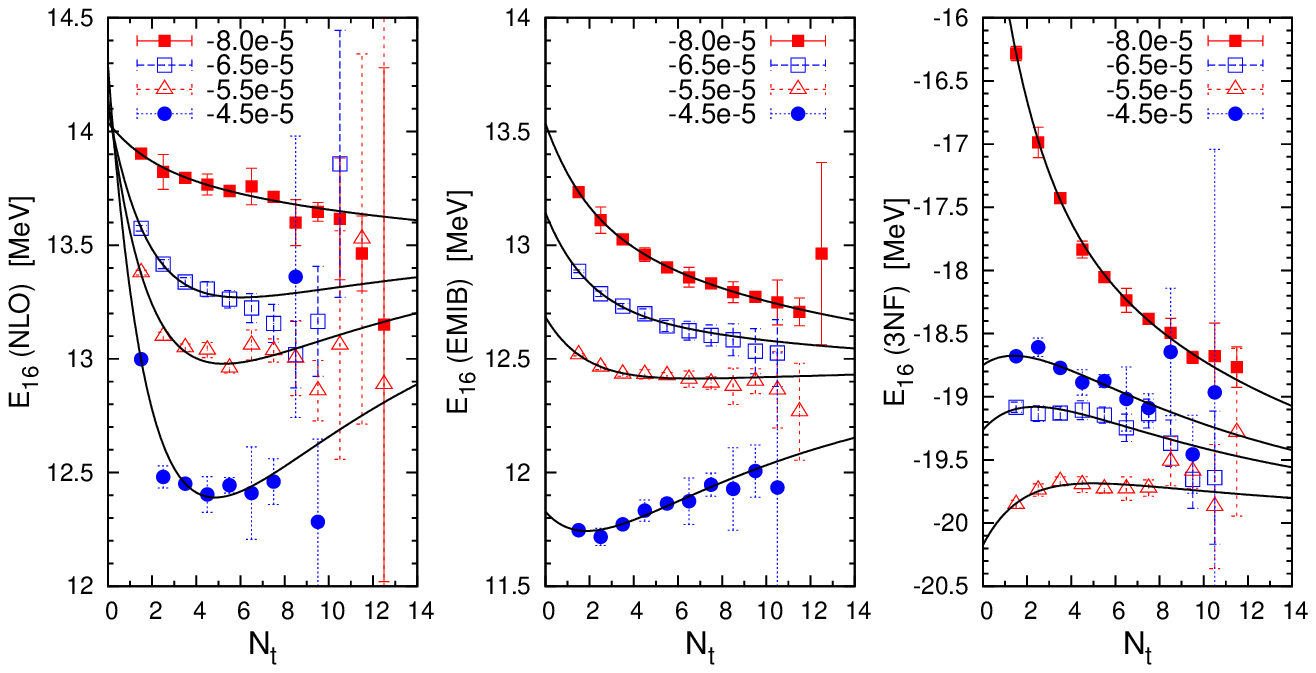}
\includegraphics[width=.7\columnwidth]{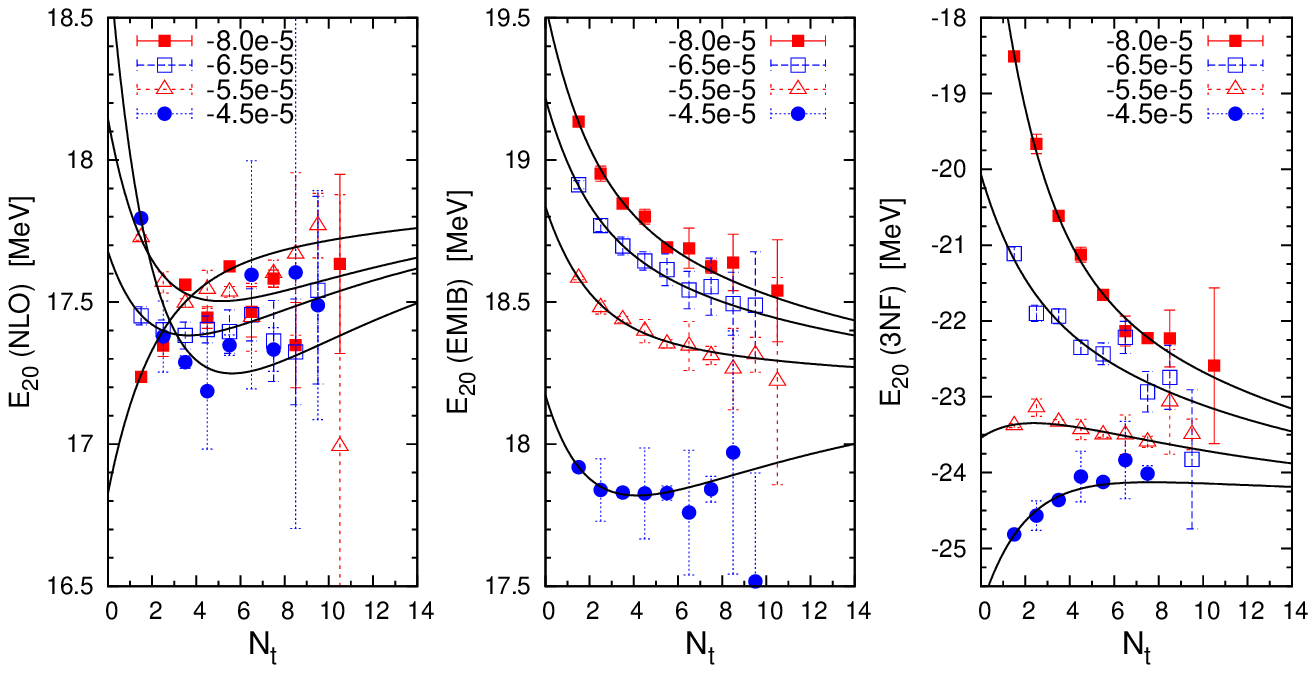}
\end{center}
\vspace{-7mm}
\caption{NLEFT results for the matrix elements $X_A^{\mathcal{O}}(t)$ for $A = 16$ and~$A = 20$, with $C_\mathrm{SU(4)}^{}$ given 
(in MeV$^{-2}$) for each trial state. The left panels show the total
isospin-symmetric NLO correction, the central panels the electromagnetic and isospin-breaking (EMIB) corrections, and the right panels the total
three-nucleon force (3NF) correction. The curves show a fit with $\rho_A^{}(E)$ given by the sum of three energy delta functions, 
correlated with those of Fig.~\ref{fig:LO}.
\label{fig:NLO1}}
%\vspace{-6mm}
\end{figure}

\begin{figure}[t]
\begin{center}
\includegraphics[width=.7\columnwidth]{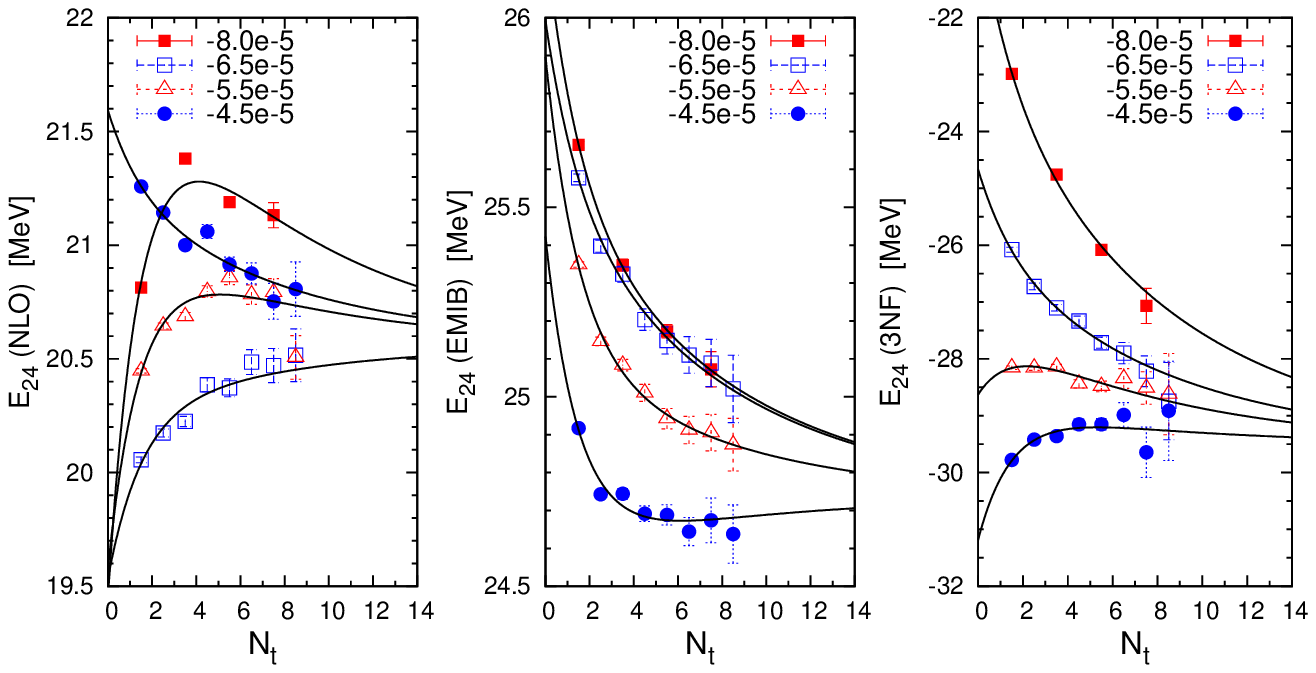}
\includegraphics[width=.7\columnwidth]{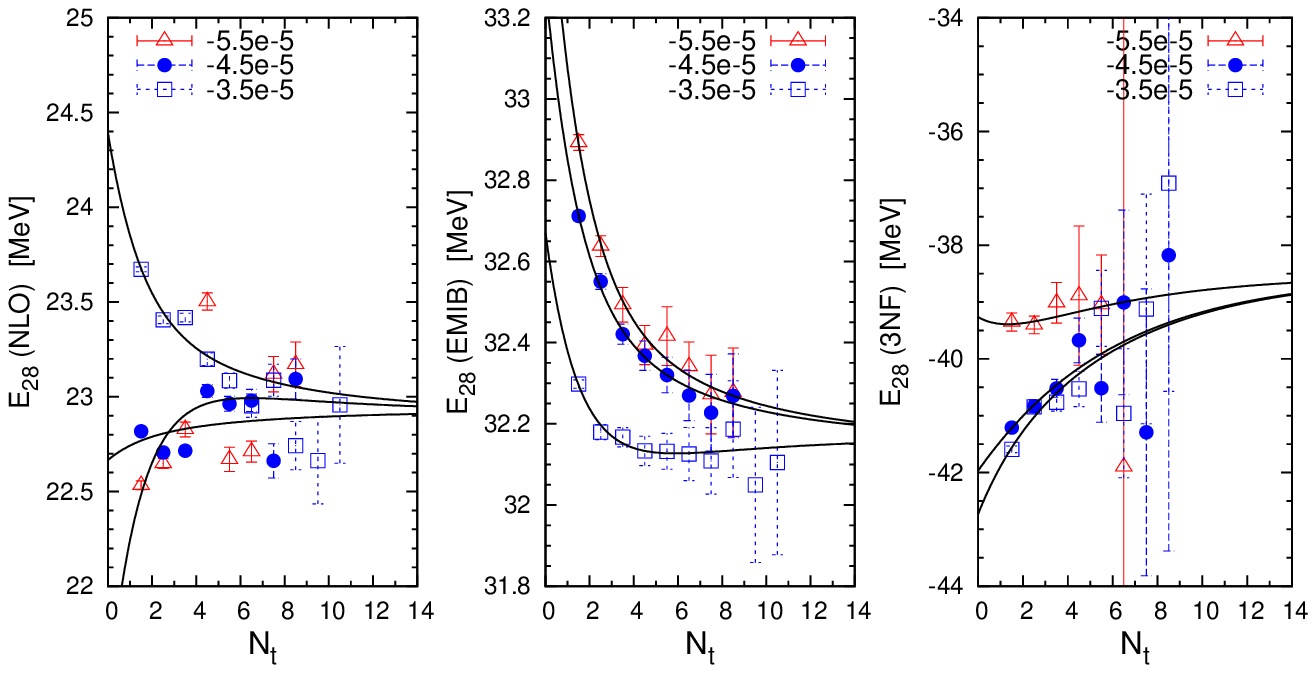}
\end{center}
\vspace{-7mm}
\caption{NLEFT results for the matrix elements $X_A^{\mathcal{O}}(t)$ for $A = 24$ and $A = 28$. The notation is
as for Fig.~\ref{fig:NLO1}. \label{fig:NLO2}}
%\vspace{-6mm}
\end{figure}

In Fig.~\ref{fig:LO}, we show the LO transient energy $E_A^{}(t)$ as a function of
the number of temporal lattice steps $N_t^{} = t/a_t^{}$, for $^{16}$O through $^{28}$Si.
The curves show a simultaneous fit to all trial states employed, with
$\rho_A^{}(E)$ given by a sum of three energy delta functions. In Figs.~\ref{fig:NLO1} and~\ref{fig:NLO2},
we similarly show the expectation values $X_A^{\mathcal{O}}(t)$ for $^{16}$O through $^{24}$Mg. These include the 
sum of isospin-symmetric NLO corrections (NLO), the sum of the electromagnetic and 
strong isospin-breaking corrections (EMIB), and the total three-nucleon force contribution~(3NF) which first appears at NNLO.
It should be noted that the fits shown in Fig.~\ref{fig:LO} are correlated with those of Figs.~\ref{fig:NLO1} and~\ref{fig:NLO2},
and use the same spectral density $\rho_A^{}(E)$. Moreover, each of the $\simeq 30$ contributions 
$X_A^{\mathcal{O}}(t)$ to the NLO, EMIB and 3NF corrections is individually accounted for in the analysis. 
We also emphasize that the fits for each $A$ are independent.

Our NLEFT results for the alpha nuclei from $^{4}$He to $^{28}$Si are summarized in Table~\ref{tab_en},
%$^{4}$He, $^{8}$Be, $^{12}$C, $^{16}$O, $^{20}$Ne, $^{24}$Mg 
with statistical and extrapolation uncertainties shown in parentheses. For comparison, we also show the empirical ground-state energies. 
The LO energies are given in the second column of Table~\ref{tab_en}, while the third column 
shows the results using the two-nucleon force up to NNLO. Our ``LO'' calculations are actually improved LO calculations with smeared 
short-range interactions that capture a significant portion of the corrections usually treated at NLO~\cite{Borasoy:2006qn}. 
The fourth column includes the 3NF at NNLO. 
As discussed in Ref.~\cite{Epelbaum:2009pd}, the local 3N contact interaction induces
significant lattice artifacts when acting on configurations of four nucleons at the same lattice site. Following 
Ref.~\cite{Epelbaum:2009pd}, we have removed this systematic effect by subtraction of a local 4N contact interaction. In the
column labeled ``+3N'' in Table~\ref{tab_en}, the strength of this subtraction has been set to reproduce the empirical binding energy of $^8$Be. 
We have not yet included systematic errors due to the finite-volume effects in a box of size $L=11.8$~fm, but preliminary results at larger volumes 
are suggestive of a $\sim 1\%$ reduction in the binding of each nucleus in the infinite-volume limit. In particular, we expect that $\sim 50$\% of the
observed $\simeq 0.7$~MeV overbinding of $^4$He should vanish.

Our NNLO results with the 3NF included appear to be within a few percent of the empirical energies for $A \leq 12$, while for 
$^{16}$O we find an overbinding of $\simeq 9\%$. Such an accuracy is, by itself, reasonably good for a calculation which is truncated at NNLO
at a lattice spacing of $a = 1.97$~fm. However, for $^{20}$Ne the observed overbinding increases to $\simeq 15$\%, for $^{24}$Mg to $\simeq 17$\%, 
and for $^{28}$Si it reaches $\simeq 30$\%.  It is thus clearly a systematical effect. In this context, we note that other \textit{ab initio} methods using 
soft potentials encounter similar problems in the description of both light and medium-mass nuclei using the same set of 
interactions~\cite{Hagen:2012fb,Jurgenson:2013yya,Roth:2011ar}. 

\begin{table}[t]
\caption{NLEFT results for the ground-state energies (in MeV).
%of the alpha-cluster nuclei. 
The combined statistical and extrapolation errors are given in parentheses. The columns labeled ``LO (2N)'' and ``NNLO (2N)'' show
the energies at each order using the two-nucleon force only. The column labeled ``+3N'' also includes the 3NF, which first appears 
at NNLO. Finally, the column ``+4N$_\mathrm{eff}$'' includes the effective 4N contribution from Eq.~(\ref{4N}). The column ``Exp'' gives
the empirical energies.
\label{tab_en}}
%\vspace{-4mm}
\begin{center}
\begin{tabular}{|r | r | r r r | r|}
\hline
%\multicolumn{1}{c}{} & \multicolumn{1}{c}{LO} & \multicolumn{3}{c}{NNLO} & \\ 
$A$ & \multicolumn{1}{c |}{LO (2N)} & \multicolumn{1}{c}{NNLO (2N)} 
& \multicolumn{1}{c}{+3N} & \multicolumn{1}{c |}{+4N$_\mathrm{eff}$} &
\multicolumn{1}{c |}{Exp} \\ \hline\hline
%$^4$He 
$4$ & $-28.87(6)$ & $-25.60(6)$ & $-28.93(7)$ & $-28.93(7)$ & $-28.30$ \\
%$^8$Be 
$8$ & $-57.9(1)$ & $-48.6(1)$ & $-56.4(2)$ & $-56.3(2)$ & $-56.35$ \\
%$^{12}$C 
$12$ & $-96.9(2)$ & $-78.7(2)$ & $-91.7(2)$ & $-90.3(2)$ & $-92.16$ \\
%$^{16}$O 
$16$ & $-147.3(5)$ & $-121.4(5)$ & $-138.8(5)$ & $-131.3(5)$ & $-127.62$ \\
%$^{20}$Ne 
$20$ & $-199.7(9)$ & $-163.6(9)$ & $-184.3(9)$ & $-165.9(9)$ & $-160.64$ \\
%$^{24}$Mg 
$24$ & $-253(2)$ & $-208(2)$ & $-232(2)$ & $-198(2)$ & $-198.26$ \\
%$^{28}$Si 
$28$ & $-330(3)$ & $-275(3)$ & $-308(3)$ & $-233(3)$ & $-236.54$\\
\hline
\end{tabular}
%\vspace{-5mm}
\end{center}
\end{table}

Before we discuss the challenge of resolving this overbinding problem in future \textit{ab initio} calculations, it is useful to explore 
the nature of the missing physics in the present work.
As we ascend the alpha ladder from $^{4}$He to $^{28}$Si, the lighter nuclei can be described as collections of 
alpha clusters~\cite{Epelbaum:2012qn,Epelbaum:2011md}. As the number of clusters increases, they become increasingly densely packed, such that
a more uniform liquid of nucleons is approached. This increase in the density of alpha clusters appears correlated with the gradual overbinding 
we observe at NNLO for $A \geq 16$. As this effect becomes noticeable for $^{16}$O, we can view it as a problem which first
arises in a system of four alpha clusters. The alpha-cluster structure of $^{16}$O will be discussed in more detail in a forthcoming 
publication~\cite{16O_spectrum}. Following Ref.~\cite{Epelbaum:2009pd}, which removed discretization errors associated with four 
nucleons occupying the same lattice site,
we can attempt to remove similar errors associated with four alpha clusters in close proximity on neighboring lattice sites. 
The simplest interaction which permits a removal of the overbinding associated with such configurations is 
%of the form
%
\begin{align}
V^{(4\mathrm{N_{eff}})}
& = D^{(4\mathrm{N_{eff}})} \hspace{-.6cm}
\sum_{1\le(\vec n_{i}^{}-\vec{n}_{j}^{})^2_{}\le2} \hspace{-.5cm}
\rho(\vec n_1^{})\rho(\vec n_2^{})\rho(\vec n_3^{})\rho(\vec n_4^{}), \label{4N}
\end{align}
with $\rho(\vec{n})$ the total nucleon density. The summation includes nearest or 
next-to-nearest neighbor (spatial) lattice sites.

In Table~\ref{tab_en}, the column
labeled ``+4N$_\mathrm{eff}$'' shows the results at NNLO while including both the 3NF and $V^{(4\mathrm{N_{eff}})}$. Due to the low momentum cutoff, the
two-pion exchange contributions have been absorbed into the contact interactions at NLO.
We have tuned $D^{(4\mathrm{N_{eff}})}$ to give approximately the correct energy for the ground state of $^{24}$Mg. With 
$V^{(4\mathrm{N_{eff}})}$ included, a good description of the ground-state energies is obtained over the full range from light to medium-mass nuclei, with
a maximum error no larger than $\sim 3$\%. This lends support to the qualitative picture that the overbinding of the NNLO results in 
Table~\ref{tab_en} is associated with the increased packing of alpha clusters and the eventual crossover to a uniform nucleon liquid.  
The missing physics would then be comprised of short-range repulsive forces that counteract the dense packing of alpha clusters. 

In spite of the good agreement with experiment in Table~\ref{tab_en} upon introduction of $V^{(4\mathrm{N_{eff}})}$, we also need to consider whether 
this could be merely an accidental effect. It is then helpful to check whether a consistent
picture is obtained with respect to excited states, transitions and electromagnetic properties of the nuclei in the medium-mass range where 
$V^{(4\mathrm{N_{eff}})}$ gives a sizable contribution. In Ref.~\cite{16O_spectrum}, we find very convincing evidence supporting our results 
and analysis from the properties of $^{16}$O, including its first excited 
$0^+$ state. In particular, the excitation energies and level ordering in $^{16}$O was found to be very sensitive to the strength and form of 
$V^{(4\mathrm{N_{eff}})}$. Such a sensitivity arises due to the differences in the alpha-cluster structure of the states in question.

The coefficient of $V^{(4\mathrm{N_{eff}})}$ can be expressed 
as $D^{(4\mathrm{N_{eff}})} = 0.9/(f^7_\pi\Lambda_\chi)$, where we use $f_\pi^{}=92.4$~MeV and  $\Lambda_\chi^{}=700$~MeV as in 
Ref.~\cite{Epelbaum:2002vt}.  
%We find that $D^{(4\mathrm{N_{eff}})}$ itself is of natural size in the chiral expansion. 
While the magnitude of $D^{(4\mathrm{N_{eff}})}$ appears to be somewhat large compared to what is expected based on 
naive dimensional analysis, the effective 4N contribution to {\it e.g.} the alpha particle binding energy is 
negligibly small, in agreement with the chiral power counting. 
Thus, the increasing effective 4N
contributions that we find for $A\geq 16$ are the result of large operator expectation values for the nuclear wave function.  
We expect that this effect is due to the coarse lattice spacing, and can be ameliorated by using a smaller lattice spacing and an interaction 
with more short-range repulsion.

\section{Conclusions}

Let us now return to the question we posed at the beginning: How large a nucleus can be calculated from first principles using 
chiral nuclear EFT, and what are the remaining challenges? Using a soft nucleon-nucleon interaction with a momentum cutoff scale 
of $\pi/a \simeq 314$~MeV, we found that the two-nucleon potential is accurate for lighter nuclei but overbinds those beyond $^{16}$O. 
As a result, the overall contribution of multi-nucleon forces must compensate by changing sign from attractive to repulsive with increasing $A$. 
While such an effect cannot be accommodated by the 3NF at NNLO alone, 
%The problem could be fixed by going to higher orders 
%in the EFT expansion, as 
the overall contribution from higher-order interactions should be similar to our effective four-nucleon interaction, which counteracts
the packing of alpha clusters. Still, this implies a large correction from higher-order terms. Analogous problems will arise in 
computational methods that use renormalization group flows to soften the two-nucleon interaction. In that case, the large repulsive corrections appear 
in the form of strong induced multi-nucleon forces.   

From our analysis, the path forward for \textit{ab initio} calculations of heavier nuclei using chiral nuclear EFT appears clear. 
The softening of the two-nucleon interaction should not be pushed so far that heavier nuclei become significantly overbound by the 
two-nucleon force alone. This is not merely an issue for NLEFT, but would appear to be a universal criterion for all \textit{ab initio} methods.  
A concerted effort should be made to improve the current computational algorithms to handle interactions with more short-range repulsion. 
The NLEFT collaboration is now exploring this approach for studies of larger nuclei. We are now in the process of improving the 
lattice algorithms for calculations at smaller lattice spacings, and extending NLEFT to N$^3$LO in the chiral expansion.

\section*{Acknowledgments}

We are grateful for the help in automated data collection by Thomas Luu.  We acknowledge partial financial support from the 
Deutsche Forschungsgemeinschaft (Sino-German CRC 110), the Helmholtz Association (Contract No.\ VH-VI-417), 
BMBF (Grant No.\ 05P12PDFTE), the U.S. Department of Energy (DE-FG02-03ER41260), and the U.S. National Science Foundation
(PHY-1307453). Further support
was provided by the EU HadronPhysics3 project and the ERC Project No.\ 259218 NUCLEAREFT. The computational resources 
were provided by the J\"{u}lich Supercomputing Centre at  Forschungszentrum J\"{u}lich and by RWTH Aachen.

%%%%%%%%%%%%%%%%%%%%%%%%%%%%%%%%%%%%%%%%%%

%\section*{References}

\end{document}